\documentclass[aps,prb,twocolumn,superscriptaddress]{revtex4-2}
\usepackage{mhchem,graphicx,longtable,xfrac}
\usepackage[colorlinks=true,citecolor=blue,linkcolor=blue,breaklinks=true]{hyperref}
\usepackage{amssymb}
\usepackage{graphicx}
\usepackage{amsmath}

\usepackage{times}
\usepackage{color}
\usepackage{bm}
\usepackage{soul}
\renewcommand{\hl}[1]{#1}

\begin{document}

\newcommand{\gvs}{GaV$_4$S$_8$}
\newcommand{\gvse}{GaV$_4$Se$_8$}

\title{Magnetoentropic mapping and computational modeling of cycloids and skyrmions in the lacunar spinels \gvs{} and \gvse{}}

\author{Julia L. Zuo}
\email[]{jlzuo@ucsb.edu}
\affiliation{Materials Department, University of California, Santa Barbara, California 93106}

\author{Daniil Kitchaev}
\email[]{dkitch@ucsb.edu}
\affiliation{Materials Department, University of California, Santa Barbara, California 93106}

\author{Emily C. Schueller}
\affiliation{Materials Department and Materials Research Laboratory, University of California, Santa Barbara, California 93106}

\author{Joshua D. Bocarsly}
\affiliation{Materials Department and Materials Research Laboratory, University of California, Santa Barbara, California 93106}

\author{Ram Seshadri}
\affiliation{Materials Department and Materials Research Laboratory, University of California, Santa Barbara, California 93106}

\author{Anton Van der Ven}
\affiliation{Materials Department, University of California, Santa Barbara, California 93106}

\author{Stephen D. Wilson}
\email[]{stephendwilson@ucsb.edu}
\affiliation{Materials Department, University of California, Santa Barbara, California 93106}

\date{\today}

\begin{abstract}
We report the feasibility of using magnetoentropic mapping for the rapid identification of magnetic cycloid and skyrmion phases in uniaxial systems, based on the \gvs{} and \gvse{} model skyrmion hosts with easy-axis and easy-plane anisotropies respectively.
We show that these measurements can be interpreted with the help of a simple numerical model for the spin Hamiltonian to yield unambiguous assignments for both single phase regions and phase boundaries.
In the two lacunar spinel chemistries, we obtain excellent agreement between the measured magnetoentropic features and a minimal spin Hamiltonian built on Heisenberg exchange, single-ion anisotropy, and anisotropic Dzyaloshinskii-Moriya interactions. 
In particular, we identify characteristic high-entropy behavior in the cycloid phase that serves as a precursor to the formation of skyrmions at elevated temperatures and is a readily-measurable signature of this phase transition.
Our results demonstrate that rapid magnetoentropic mapping guided by numerical modeling is an effective means of understanding the complex magnetic phase diagrams innate to skyrmion hosts. 
One notable exception is the observation of an anomalous, low-temperature high-entropy state in the easy-plane system \gvse, which is not captured in the numerical model. 
Possible origins of this state are discussed.
\end{abstract}

\maketitle

\section{Introduction}
The development of next-generation spintronic devices relies in a large part on engineering subtle magnetic phase transitions which control the formation of long-wavelength spin textures. 
In particular, topologically non-trivial spin textures such as skyrmion lattices are of interest for high-density, energy-efficient non-volatile magnetic memory schemes \cite{appl1, appl2}. 
While a number of materials hosting skyrmion phases have been reported, the characterization of their inherently complex phase diagrams and the bounds of their thermodynamic stability remains a daunting experimental task reliant on multimodal investigation. 
Skyrmion states are typically identified using small-angle neutron scattering (SANS) and Lorenz-mode transmission electron microscopy (LTEM)\cite{adams2012,phatak2016,yu2010,tokunaga2015}.
Both techniques impose substantial limitations on sample geometries and are not amenable to high-throughput investigation. 
As in other areas of functional materials development \cite{Brandt2017}, the identification of practical skyrmion host materials requires much faster, low-cost techniques for magnetic phase mapping so as to allow for an expansive study of candidate materials.

One candidate rapid characterization technique for mapping complex magnetic phase diagrams is magnetoentropic mapping using rapid magnetization measurements taken across an array of applied fields and temperatures. 
Magnetoentropic mapping uses the Maxwell relation between magnetization and entropy to resolve the entropic susceptibility $(dS/dH)_T$, which can in principle unambiguously distinguish skyrmion phases from long wavelength spin modulations and other magnetic phases \cite{joshFeGe}. 
This technique is complementary to conventional magnetic susceptibility $(dM/dH)_T$ measurements that can identify magnetic phase boundaries \cite{Pfleiderer2010,bordacs,loidlgvs,Butykai2017} but are unable to distinguish skyrmions from other metamagnetic phase transitions. 
Several proof-of-concept studies have shown that magnetoentropic mapping can identify skyrmion stability regions near $T_c$ (the Curie temperature) in conventional cubic skyrmion hosts \cite{joshFeGe, Kautzsch2020, Ge2015, Han2017, Dhital2020}.
However, the applicability of this technique to more general types of magnetic phase diagrams remains undemonstrated \cite{Jama2019}.
In particular, the magnetoentropic behavior of uniaxial systems remains unresolved, which includes both materials with bulk uniaxial symmetry and thin film systems where skyrmions may persist over wide temperature windows far below $T_c$ \cite{kitchaevmodel,Banarjee2014,Gungordu2016}.

\begin{figure}
\includegraphics[scale=1]{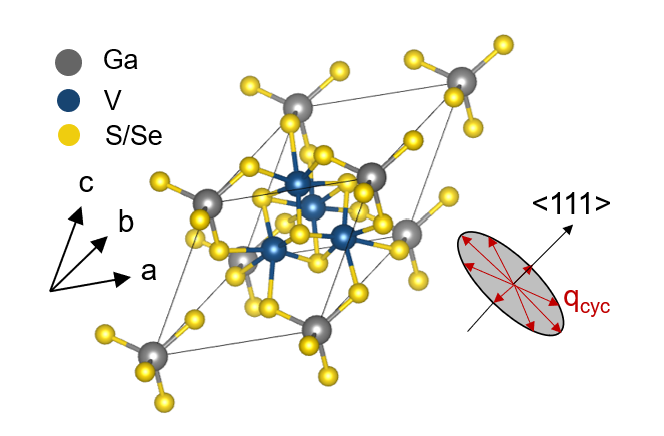}%
\caption{\label{fig:structurefig} At T$_S$ = 42 K, \gvs{} and \gvse{} distort rhombohedrally to $R3m$, shown above. This ferroelectic transition allows cycloidal modulations with propogation vectors in the plane perpendicular to $\langle 111 \rangle$.}
\end{figure}

A class of model uniaxial skyrmion hosts that exemplifies the rich topology of skyrmionic phase diagrams is the lacunar spinel family of structures. 
These compounds are defined by tetrahedral clusters of transition metals with a collective spin state that undergo a distortion from $F\bar{4}3m$ to $R3m$ or $Imm2$ symmetry upon cooling \cite{Francois1992,pocha2000,Bichler2008}.
This structural symmetry breaking leads to the emergence of magnetic Dzyaloshinskii-Moriya interactions.
These interactions perturb the nominal ferromagnetic order by stabilizing long-wavelength modulations that lead to cycloid and skyrmion states.
Depending on the strength of the uniaxial anisotropy, these non-collinear states may appear across an extended range of temperatures or only near $T_c$ \cite{kitchaevmodel}.

In all cases, cycloidal propagation vectors lie in the plane perpendicular to the $\langle 111 \rangle$ direction of the high-$T$ cubic structure shown in Figure \ref{fig:structurefig} (equivalently, perpendicular to the $\langle 0 0 1 \rangle$ direction in the low-$T$ $R3m$ phase). 
This behavior, including the emergence of N\'eel skyrmion lattices, has been demonstrated in \gvs{}   \cite{loidlgvs,Ruff2015,White2018}, \gvse{} \cite{bordacs,ruffgvse,fujima,schuellergvse}, GaMo${}_4$S${}_8$ \cite{kitchaevmodel, Butykai2019, Zhang2019}, and GaMo${}_4$Se${}_8$ \cite{schuellergmose}.
The magnetic phase boundaries of these materials have been mapped out by conventional techniques, including SANS, field-dependent magnetization, probe microscopy, and AC susceptibility \cite{bordacs, loidlgvs, schuellergmose,Geirhos2020,Gross2020}.
Furthermore, lacunar spinels are ferroelectric \cite{bordacs, afm, loidlgvs} and susceptible to structural phase transitions \cite{schuellergmose}, meaning that skyrmion stability can be probed using electric fields and mechanical strain \cite{Geirhos2020}.
Finally, while the electronic structure of these materials is highly correlated and requires advanced computational methods to resolve \cite{schuellergvse, Vanderbilt2020}, their microscopic magnetic behavior is well-captured using semi-classical spin models \cite{kitchaevmodel, Zhang2019, nikolaev2020}.
This combination of rich physics, abundant experimental data and  quantitative theory establish the lacunar spinels as an ideal system for testing new methods for magnetic phase determination.

Here, we use high-throughput magnetocaloric measurements to perform magnetoentropic phase mapping in \gvs{}  ~and \gvse{} , two uniaxial lacunar spinel skyrmion hosts with drastically different magnetic phase diagrams.
We characterize the magnetic entropy of both materials experimentally and with Monte Carlo simulations and then use these data to unambiguously label nearly all phases and phase boundaries.
We identify characteristic positive-entropy precursor effects in the cycloid phase near the cycloid-to-skyrmion transition, which can be used to identify the phase boundary between these two states.
Furthermore, we identify an unusual high entropy region at the low temperature, high-field phase boundary of the cycloidal state in \gvse{} , which may be a signature of recently proposed magnetic phases confined to polar domain walls \cite{Geirhos2020}.
More generally, our results demonstrate that magnetoentropic mapping informed by computational models of entropic susceptibility can provide a rapid, unambiguous measurement of cycloid/skyrmion phase boundaries in model uniaxial systems and thereby facilitate the rapid characterization of similar skyrmion phase diagrams.

\section{Methods}

Phase pure polycrystalline samples were synthesized from stoichiometric amounts of elemental Ga, V, and S or Se. 
Reagents were sealed in fused silica under vacuum and reacted in a box furnace at 950\,$^{\circ}$C for 72 hours with a 1\,$^{\circ}$C/min ramp on heating and furnace cooled. 

Single crystals of \gvs\ and \gvse\ were grown using the chemical vapor transport method. Polycrystalline source powders were loaded in 11 inch long evacuated fused silica tubes with \ce{I_2} (10 mg/cm$^3$) or \ce{PtCl_2} transport agents for \gvs\ or \ce{GaV_4Se_8,} respectively. 
\gvs\ crystals were grown for two weeks with the powder source at 850\,$^{\circ}$C and the growth zone at 800\,$^{\circ}$C. 
\gvse\ crystals were grown for two weeks with the powder source at 960\,$^{\circ}$C and the growth zone at 920\,$^{\circ}$C. 

Magnetic measurements were performed on a Quantum Design Magnetic Property Measurement System 3. 
Single crystals were mounted on low-background quartz paddles using GE-7031 varnish with the desired crystal orientation along the direction of the applied magnetic field. 
Magnetoentropic phase maps were determined by measurement of field cooled magnetization \textit{versus} temperature at uniform external field intervals in the temperature range of interest.

Magnetization versus temperature data was smoothed before taking the derivative using the magentro.py python script (Figure \ref{fig:F1}) \cite{joshFeGe}.
A detailed description of the method used and the open source code can be found in the original manuscript \cite{joshFeGe}.

Monte Carlo sampling was performed using an in-house implementation of magnetic Hamiltonian Monte Carlo \cite{Wang2019}, based on a generalized formulation of spin-spin interactions described in previous works \cite{kitchaevmodel, schuellergmose}.
Briefly, each tetrahedral V${}_4$ cluster is assumed to have a collective moment equal to 1 $\mu_B$, \cite{kim2014,schuellergvse} which is allowed to interact quasi-classically with nearest-neighbor tetrahedra through a combination of Heisenberg exchange and symmetry-restricted anisotropic Dzyaloshinskii-Moriya interactions \cite{Kitchaev2018,avdv2018}. 
Additionally, each V${}_4$ moment is assigned a uniaxial, single-ion anisotropy energy.
This model is derived by considering all possible interactions on this magnetic sublattice, and eliminating terms that are found to be insignificant or redundant in known lacunar spinel skyrmion hosts \cite{kitchaevmodel}.
Thermodynamic data was obtained using 10000 independent samples following an initial equilibration run of 1000 independent samples, where the number of Monte Carlo iterations between samples was automatically set by the decay constant of the energy autocorrelation function.
All Monte Carlo runs were performed using a three step protocol: (1) zero-$T$ ground state optimization using simulated annealing and conjugate gradient descent starting from known spin textures, (2) constant-field heating from 1 K to above the Curie temperature, (3) constant-field cooling back to 1 K.
Phase boundaries were then identified on the basis of the magnetic structure factor, discontinuities in magnetization, susceptibility and heat capacity, and the topological index.
The topological index is computed using the expression
\[
t = \bigg \langle \frac{1}{4\pi} \int m \cdot \left(\partial_x m \times \partial_y m \right)\mathrm{d}x\mathrm{d}y \bigg \rangle
\]
where $m$ is the local magnetization, the expectation is taken over all sampled microstates, and $z$ is taken to be the skyrmion normal axis \cite{Hoffmann2017}.

\section{Results}

\begin{figure}
\includegraphics[scale=0.75]{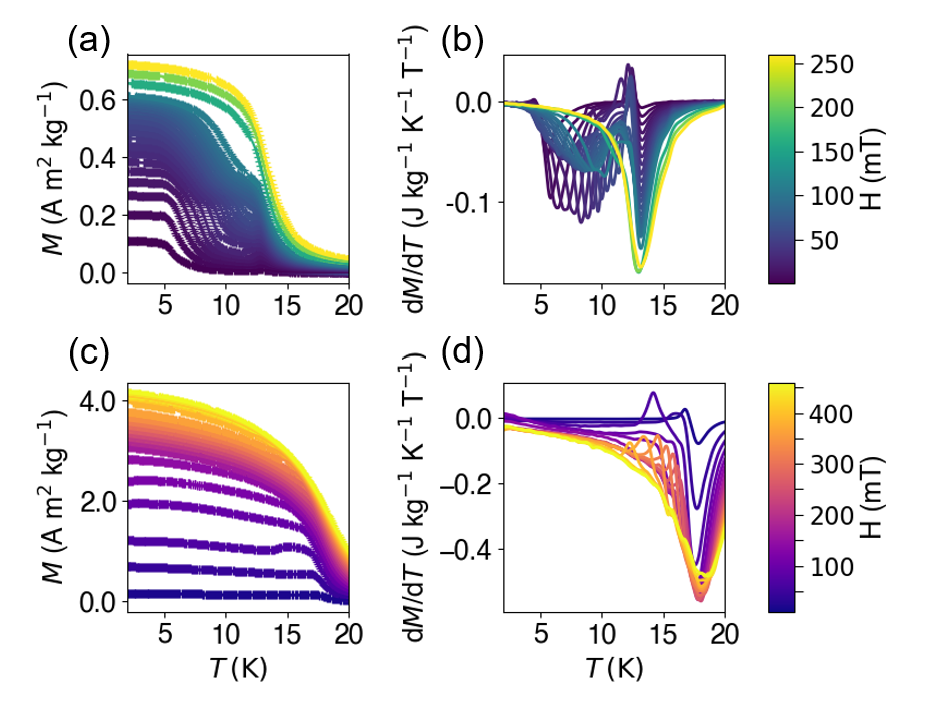}%
\caption{\label{fig:F1} Constant-field magnetoentropic measurements of \gvs{} and \gvse. The magnetization $M$ versus temperature $T$ is measured at even intervals of external field for \gvs{} and \gvse{} as shown in panels (a) and (c) respectively. Data points (+) are extremely dense, so a statistical smoothing routine is applied (solid lines) before taking the derivative to obtain the derivatives $(dM/dT)_H = (dS/dH)_T$ as shown for \gvs{} and \gvse{} in panels (b) and (d) respectively.}
\end{figure}

The foundation of magnetoentropic mapping is the thermodynamic equivalence between the evolution of magnetization with temperature and entropy with field within a single phase, given by the Maxwell relation 
\[
\left. \frac{dM}{dT} \right|_{H} = \left. \frac{dS}{dH} \right|_{T}
\] 
This relationship can be integrated to determine the entropy change with field within a given phase.
At first-order phase transitions, magnetization and entropy change discontinuously, as given by the Clausius-Clapeyron relation 
\[
\Delta S_{\alpha \rightarrow \beta} = -\Delta M_{\alpha \rightarrow \beta} \frac{dH_{\alpha \rightarrow \beta}}{dT_{\alpha \rightarrow \beta}}
\]
If the geometry of phase boundaries can be inferred from a combination of features in $dM/dT$ and $dM/dH$, this equation can be used to determine the isothermal change in entropy under magnetic field across the phase diagram, $\Delta S = S(H,T) - S(0,T)$:
\[
\Delta S(H) = \int_0^{H_{\alpha \rightarrow \beta}} \frac{dS}{dH^\dagger} dH^\dagger + \Delta S_{\alpha \rightarrow \beta} + \int_{H_{\alpha \rightarrow \beta}}^{H} \frac{dS}{dH^\dagger} dH^\dagger
\]
where we use the Maxwell relationship to compute the entropy within the $\alpha$ and $\beta$ phases, and the Clausius-Clapeyron relation to determine the entropy change across the $\alpha \rightarrow \beta$ phase transition. 
Following these relations, magnetic entropy can be estimated from pure magnetization data \cite{Amaral2010}.

However, precisely determining $\Delta S_{\alpha \rightarrow \beta}$ from the Clausius-Clapeyron relation is difficult because of the ubiquitous uncertainty in $\Delta M_{\alpha \rightarrow \beta}$.
The exact entropy change at a phase boundary can be obtained from calorimetric measurements taking into account hysteresis effects and domain phenomena \cite{Bauer2013, Amaral2009, Carvalho2011,Balli2009}, but such data is rarely available.
Fortunately, the qualitative behavior of $\Delta S_{\alpha \rightarrow \beta}$ can be inferred directly from $(dM/dT)_H$ data, where sharp peaks and valleys denote phase transitions into higher or lower entropy phases respectively \cite{joshFeGe}.
Thus, magnetoentropic mapping alone can be used to identify field-driven phase transitions in analogy to heat capacity mapping that is frequently used to trace out temperature-driven phase transitions. 
Furthermore, the relationship between $(dM/dT)_H$ and $(dS/dH)_T$ within single-phase regions means that this data can provide unique insight into the properties of the magnetic phases being observed.
We will show that in combination with numerical simulations, this data can be interpreted to obtain unambiguous signatures of the types of magnetic order that may form across the phase diagram.

The source data for magnetoentropic mapping are constant-field magnetization measurements collected at variable temperature.
A sample of this data for \gvs\ and \gvse\ is shown in Figure \ref{fig:F1}, highlighting the distinct metamagnetic features in the magnetization curves, where skyrmion formation has been reported.
The sharp changes in magnetization with field indicate that these transitions are likely to be first order.
The derivative of this data corresponds to the entropic susceptibility $(dS/dH)_T$ and shows distinct regions at various fields, suggesting that the magnetic phases observed in this system can be distinguished by their entropic behavior. 

\begin{figure*}
\includegraphics[width=\textwidth]{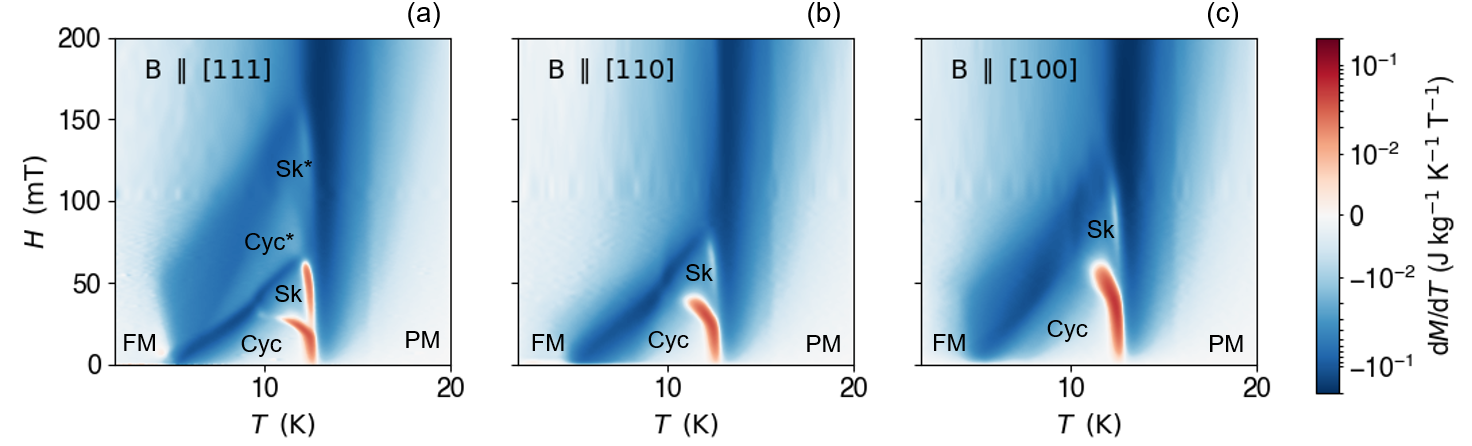}%
\caption{\label{fig:F2} \hl{Skyrmion and cycloid phase boundaries in the $(dM/dT)_H$ phase diagrams of \gvs .} When plotted as a heatmap against applied magnetic field and temperature, the skyrmion (Sk) and cycloidal (Cyc) phases are clearly visible in $dM/dT$ in several crystallographic directions, $\langle 111 \rangle$ (a), $\langle 110 \rangle$ (b), and $\langle 100 \rangle$ (c). Starred (*) phases correspond to phases in equivalent $\langle 111 \rangle$ domains that are not completely aligned with the field.}
\end{figure*}

A complete magnetoentropic map for \gvs\ is shown in Figure \ref{fig:F2}, for fields oriented along the high-symmetry directions of the crystal.
The features observable in $(dM/dT)_H$ qualitatively coincide with phase boundaries previously determined by SANS and magnetic susceptibility measurements, and are marked following the convention introduced in previous studies \cite{fujima,Ruff2015,loidlgvs}.
The cycloid and skyrmion phases in grains whose high-symmetry $c$-axis is aligned with the magnetic field direction are marked Cyc and Sk respectively.
However, in this situation the magnetic field is also oriented along inequivalent axes of structural twins within the crystal.
This superimposes magnetic phase regions from the twins with different $H$ and $T$ phase boundaries due to the smaller component of the magnetic field projected along their $\langle 111 \rangle$ directions.
These phase regions are marked in Figure \ref{fig:F2}a as Sk* and Cyc*. Because the component of the external magnetic field is smaller in the other $\langle 111 \rangle$ directions, the phase regions extend to higher field limits than the boundaries associated with the primary $\langle 111 \rangle$ domain aligned with the external field \cite{loidlgvs,tiltedBtheory}.

\begin{figure*}
\includegraphics[width=\textwidth]{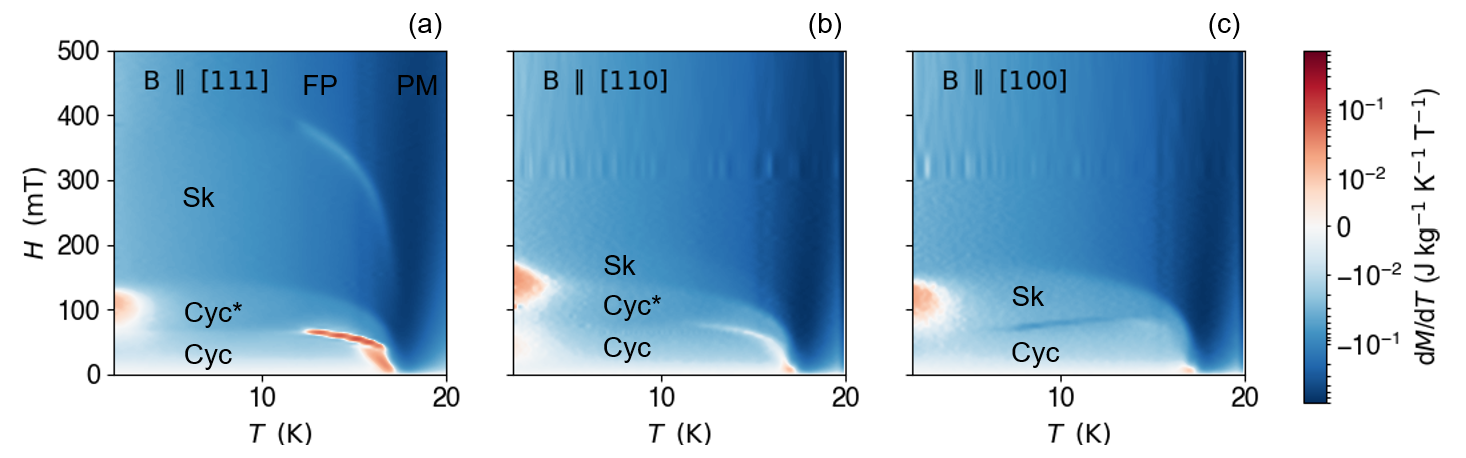}%
\caption{\label{fig:F3} Large skyrmion and cycloid phase boundaries are apparent in the $(dM/dT)_H$ phase diagrams of \gvse\ in different orientations, $\langle 111 \rangle$ (a), $\langle 110 \rangle$ (b), and $\langle 100 \rangle$ (c). Starred (*) phases again correspond to phases in equivalent $\langle 111 \rangle$ domains that are not completely aligned with the field and dashed lines correspond to transitions attributed to these domains.}
\end{figure*}

The magnetoentropic analysis of the related \gvse\ compound is shown in Figure \ref{fig:F3}.
Similarly to \gvs, features in the magnetoentropic map have the same geometry as phase boundaries previously determined by SANS and susceptibility data.
The difference in the \gvs\ and \gvse\ phase diagrams arises from the difference in their magnetocrystalline anisotropy \cite{kitchaevmodel}: \gvs\ is an \hl{easy-axis material}, favoring spin orientation along the high-symmetry $c$-axis of the $R3m$ unit cell of the low-$T$ structure ($\langle 111 \rangle$ axis of high-$T$ structure shown in Figure \ref{fig:structurefig}), while \gvse\ is an easy-plane material.
In \gvse, the skyrmion phase is destabilized in off-axis domains because the easy-plane anisotropy of the material substantially lowers the energy of the competing field-polarized phase.
In \gvs, the easy-axis anisotropy destablizes off-axis field-polarized structures and results in a much wider region of skyrmion stability against transverse applied fields.
However, the easy-plane anisotropy in \gvse\ improves skyrmion stability in temperature, allowing for these structures to persist far below the Curie temperature in the domains aligned with the applied field \cite{bordacs,kitchaevmodel}.
As a result of this change in magnetocrystalline anisotropy, the off-axis domains responsible for the Cyc* and Sk* regions in \gvs\ manifest less clearly in \gvse, but the Sk region extends to the lowest measured temperatures.

A unique feature of the \gvse{} \ magnetoentropic map is the region of positive $dS/dH$ at low temperature.
This entropic anomaly is not explained by the semi-classical spin-Hamiltonian relevant to the bulk $R3m$ phase \cite{kitchaevmodel}.
This region does coincide with previously undetermined regions of the \gvse\ phase diagram \cite{bordacs}, which were proposed to arise due to distortions of cycloidal structure within limited structural domains.
A more recent analysis of \gvse{} \ has suggested that new, distinct magnetic phases may form at these fields and temperatures, confined to structural domain walls from the $F\bar{4}3m \rightarrow R3m$ structural transition \cite{Geirhos2020}.
These microstructural mechanisms may give rise to the entropic anomaly we measure, but we are not able to definitively resolve the mechanism responsible.

\begin{figure}
\includegraphics[width=0.5\textwidth]{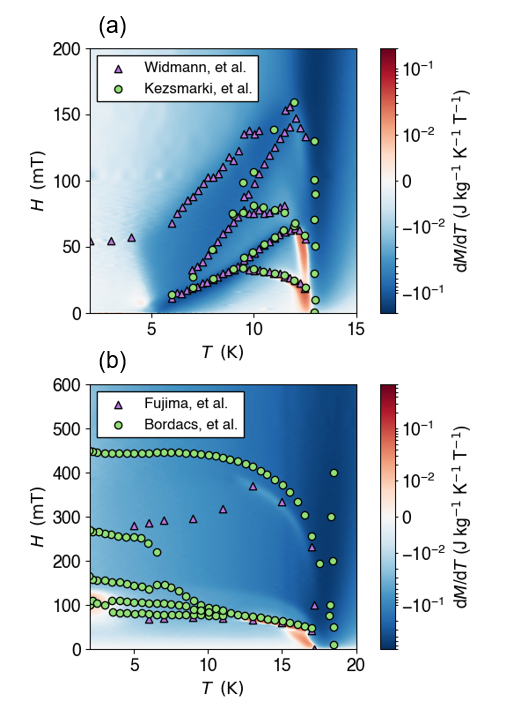}
\caption{\label{fig:F8} Magnetoentropic phase diagrams measured along the $\langle 111 \rangle$ direction compared with previously reported phase boundaries determined by other techniques for (a) \gvs\ and (b) \gvse{} }
\end{figure}

The features in the magnetoentropic data shown in Figures \ref{fig:F2} and \ref{fig:F3} correlate with with previously-determined phase boundaries \cite{widmann, loidlgvs, fujima, bordacs} as shown in Figure \ref{fig:F8}, and reveal evidence of characteristic precursor phenomena in the cycloid phase.
Most features in the magnetoentropic map coincide with known phase boundaries and can be taken as signatures of phase transitions.
However, the prominent regions of $(dM/dT)_H > 0$ near $T_c$ occur immediately below the established cycloid/skyrmion phase boundary, within the cycloid region.
Curiously, this positive $(dM/dT)_H$ feature is absent at lower temperatures in \gvse{} , even though the cycloid/skyrmion phase boundary extends from T${}_c$ down to 0K.
We will show that this prominent $dM/dT > 0$ feature arises from high-$T$  precursor phenomena in the cycloid phase that is characteristic of the cycloid-to-skyrmion transition.
This observation reflects a unique advantage of the magnetoentropic mapping technique---namely the possibility that phases may be uniquely identified by their entropic behavior.

To quantify the features in the entropic susceptibility, and their correspondence to the magnetic phases in \gvs\ and \gvse, we turn to a minimal numerical model capable of representing the bulk magnetic phase diagrams observed in the lacunar spinels.
A previous analysis of magnetic interactions in the lacunar spinels showed that magnetic phase behavior in these materials can be reduced to an effective exchange interaction, in-plane Dzyaloshinskii-Moriya coupling and single-ion anisotropy \cite{kitchaevmodel}. 
While other interactions are certainly present and necessary for reproducing the full energy landscape, equilibrium properties can be described by a three-term Hamiltonian:
\[
E = K_u \sum S_z^2 + J \sum_{i < j \in \text{NN}} S_i \cdot S_j + D \sum_{i < j \in \text{IP}} w_{ij}
\]
where $K_u$ is the single-ion uniaxial anisotropy, $J$ is the exchange constant, $D$ and $w_{ij}$ are the strength and mathematical form of the Dzyaloshinskii-Moriya interaction respectively\cite{Kitchaev2018}.
The magnetic sublattice that is being summed over by $i,j$ consists of tetrahedral V${}_4$ units, each possessing a collective moment equal to 1$\mu_B$ \cite{kim2014, schuellergvse}.
The summation over NN refers to a summation over all nearest-neighbor sites, while the summation over IP refers to a summation over nearest neighbor sites lying in the ($ab$)-plane of the $R3m$ low-$T$ structure shared by \gvs\ and \gvse.
\hl{We neglect higher-order anisotropies as past experimental and computational studies of these materials have not indicated any substantial anisotropy beyond the lowest-order term.}\cite{Kitchaev2018,loidlgvs,bordacs}
This Hamiltonian leads to cycloidal ground states at low $T$ and field, with wavenumber $q \propto D/J$ \cite{Bogdanov1999}.
The characteristic energy scale of the cycloid with respect to the ferromagnetic state is $H_\text{c} = 2\pi^2 q^2 J$.
This energy scale allows us to normalize the Hamiltonian and define a characteristic magnetization $M^*$, field along the $c$-axis $H^*$, temperature $T^*$, and uniaxial anisotropy $K_u^*$.
\begin{align*}
    M^* &= M / M_s \\
    H^* &= g\mu_{\text{B}}M_s H / H_\text{c} \\
    T^* &= k_{\text{B}} T / \sqrt{2}J \\
    K^*_u &= K / H_\text{c}
\end{align*}
where $M_s$ is the saturation magnetization \cite{kitchaevmodel}.
For \gvs{} and \gvse{}, the approximate parametrization of this Hamiltonian is given by $q=1/26$ \cite{loidlgvs} and $q=1/27$ \cite{ruffgvse}, $J = 1.4$ meV and $J = 2.0$ meV, $K^*=-1.0$ and $K^*=0.33$ respectively, based on a comparison of the experimental phase diagrams shown in Figure \ref{fig:F8} and Monte Carlo results elaborated below.

\begin{figure}
\includegraphics[width=0.5\textwidth]{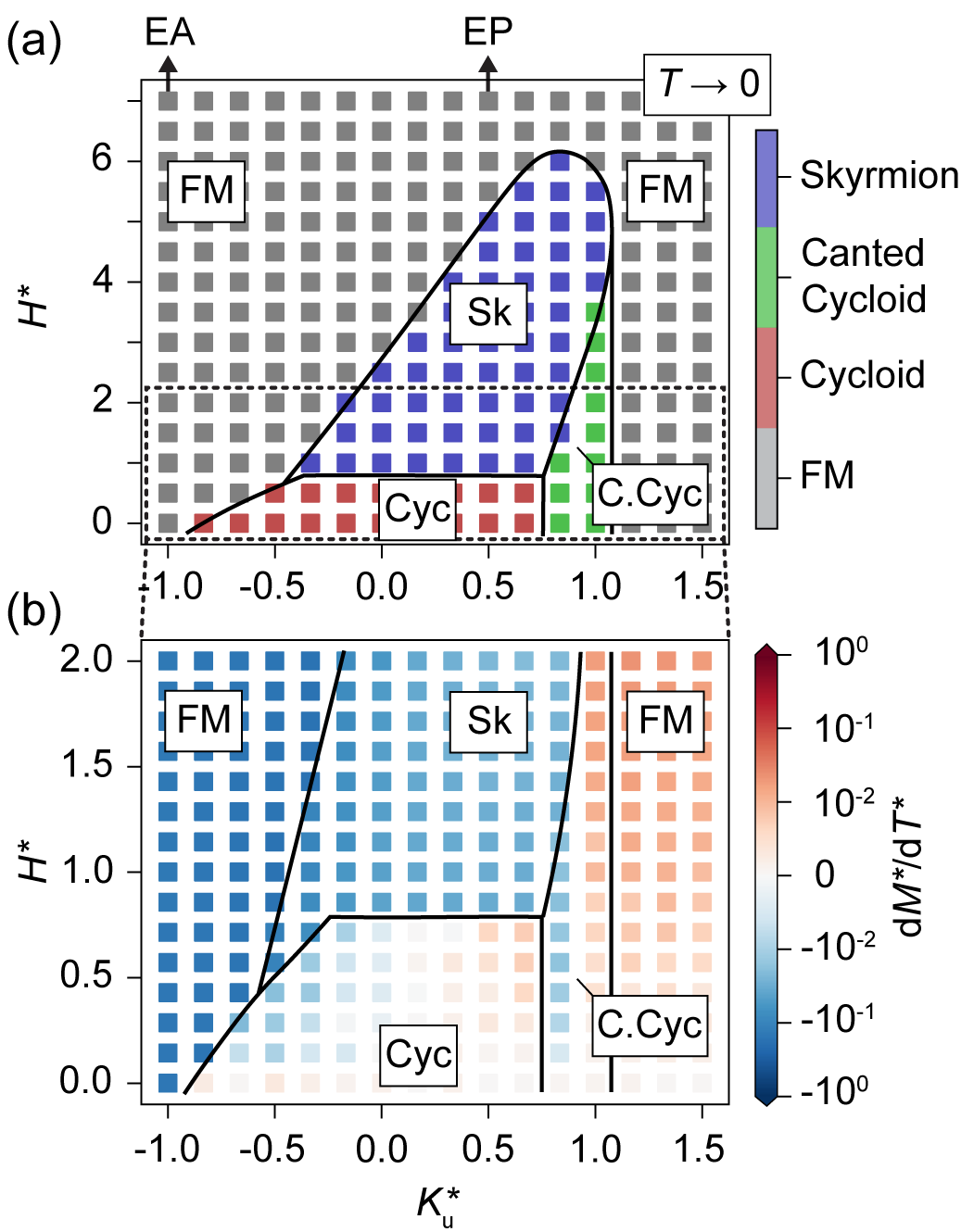}%
\caption{\label{fig:F4} Magnetic phase stability and entropic susceptibility $dS^*/dH^* = dM^*/dT^*$ in lacunar spinels in the low-$T$ limit, across possible values of uniaxial anisotropy $K_u^*$}
\end{figure}

The low-$T$ behavior of this Hamiltonian is shown in Figure \ref{fig:F4}, as a function of the applied magnetic field $H^*$ and uniaxial anisotropy $K_u^*$.
Figure \ref{fig:F4}a maps out the equilibrium phases across this space, showing clear regions of cycloid and skyrmion stability around the isotropic limit $K_u^* = 0$.
Consistent with previous reports \cite{Banarjee2014,Rowland2016,kitchaevmodel}, easy-plane anisotropy ($K_u^* > 0$) is more conducive to skyrmion formation at low $T$ than easy axis anisotropy ($K_u^* < 0$).
~\gvs{} is known to have easy-axis anisotropy, falling near the region marked EA in Figure \ref{fig:F4}a, while \gvse{} is an easy-plane material falling in the region marked EP.
Although the true anisotropy values for these two materials are difficult to compute due to the highly correlated electronic structure of both materials \cite{schuellergvse}, the conditions marked EA and EP capture the essential behavior of these two chemistries.
It must be noted that the phase diagram shown in Figure \ref{fig:F4} is constrained by periodic boundary conditions commensurate with the equilibrium cycloid wavenumber $q$.
While this approximation is valid at low fields, the high-field skyrmion-to-ferromagnet transition proceeds by a continuous increase in the skyrmion lattice constant, eventually leading to the formation of isolated skyrmions \cite{Butenko2010}.
Thus, the large discontinuity in magnetization at the high-field phase transition between a skyrmion lattice and a field-polarized ferromagnet is an artefact of the periodic boundary conditions.
As a result, we restrict our analysis primarily to low-field behavior where the periodic boundary condition approximation is appropriate.

\begin{figure*}[t!]
\includegraphics[width=\textwidth]{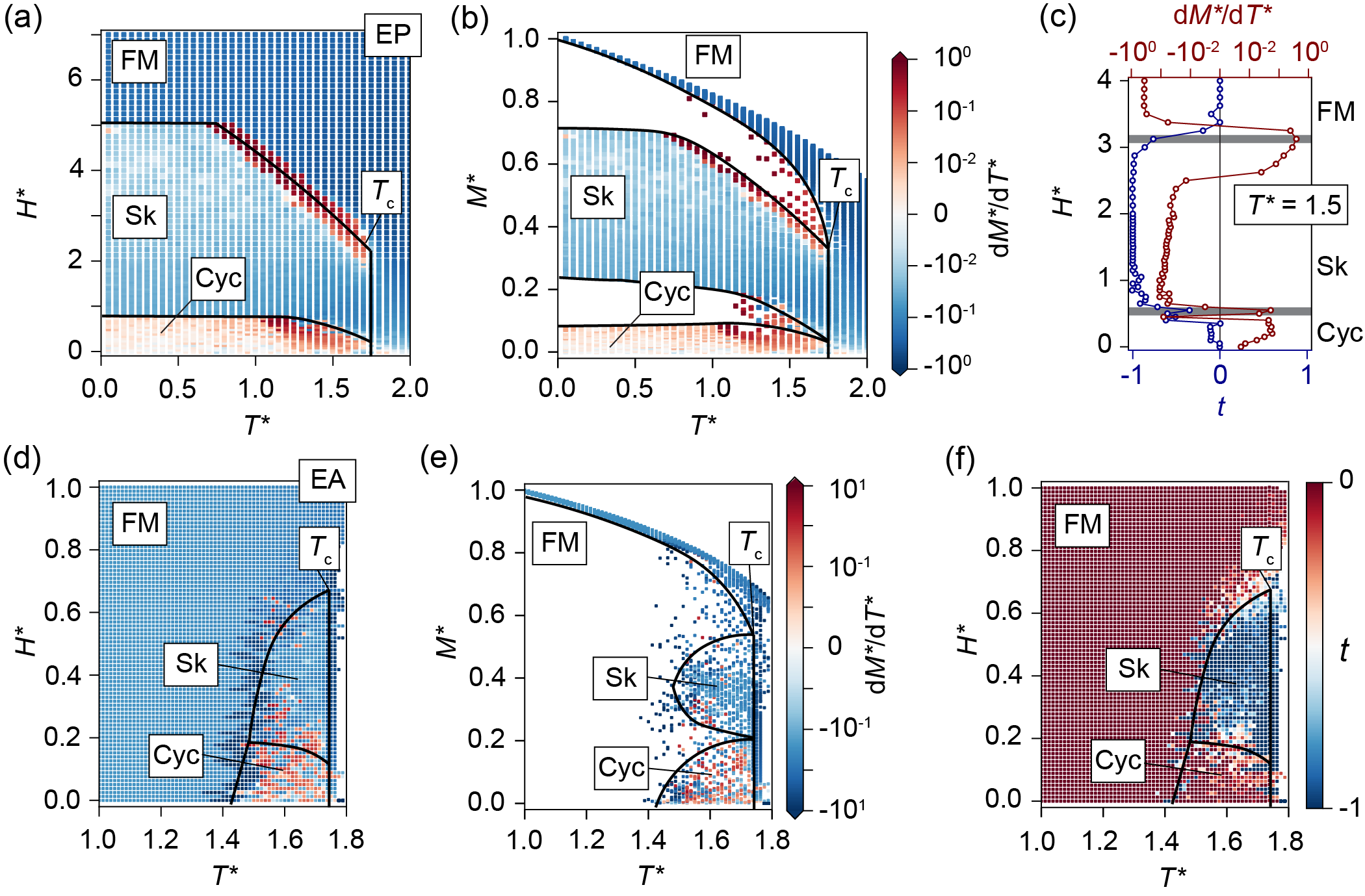}
\caption{\label{fig:F5} Computed magnetic phase diagram of an easy-plane (a-c) and an easy-axis (c-f) lacunar spinel ferromagnet, representative of \gvse{} and \gvs{} respectively. The magnetic field $H$ is oriented along the high-symmetry $c$-axis of the low-$T$ $R3m$ structure ($\langle 111 \rangle$ axis in Figure \ref{fig:structurefig}) (a, b, d, e) Phase stability as a function of normalized temperature $T^*$ and applied field or resulting magnetization respectively. The color map denotes the entropic susceptibility $dS^*/dH^* = dM^*/dT^*$ obtained from constant-field heating Monte Carlo runs. Phase boundaries are drawn on the basis of magnetization, magnetic structure factor, the topological index ($t$) and heat capacity at each point. (c,f) Correlation between $dM/dT$ and the topological index as a function of field. In the easy-plane case (c), the data is given only at $T^* = 1.5$ for clarity, with phase boundaries marked in gray.}
\end{figure*}

By comparing these phase boundaries to the magnetoentropic behavior of the Hamiltonian, we establish which phase boundaries may be unambiguously identified experimentally from $dM^*/dT^*$ measurements. 
Figure \ref{fig:F4}b shows the low-$T$ limit of $dM^*/dT^*$ across all values of anisotropy and field.
This data reveals that while $dM^*/dT^*$ does not map out all phase boundaries at low $T$, it may identify the cycloid-to-skyrmion transition in the easy-plane regime.
In all cases, skyrmions have $dM^*/dT^* < 0$, while cycloids switch from $dM^*/dT^* < 0$ to $dM^*/dT^* > 0$ near $K_u^* = 0$.
This behavior can be qualitatively explained by a switch from anisotropy favoring the same spin orientation as the magnetic field ($K_u^* < 0$) to anisotropy competing with the magnetic field ($K_u^* > 0$)

To directly assess the relationship between $dM^*/dT^*$ measurements and phase boundaries in \gvs\ and \gvse, we compute the $H^*$-$T^*$ phase diagram for EA and EP conditions respectively.  The computed phase diagram for EP conditions, representative of \gvse{} , is shown in Figure \ref{fig:F5}ab, in terms of applied field $H^*$ and resulting magnetization $M^*$ respectively.
Consistent with the experimentally-derived phase diagram of \gvse{} skyrmions are stable at all temperatures below $T_c$ at intermediate fields.
At low fields, cycloids are stable at all temperatures.
The cycloid to skyrmion and skyrmion to ferromagnet transitions appear to be first order in our simulations.
Without the restriction of periodic boundary conditions, the skyrmion to ferromagnet transition at high field may become more continuous, but to the best of our knowledge the cycloid to ferromagnet transition at low-field is correctly captured as a first-order boundary \cite{Butenko2010}.

The evolution of $dM^*/dT^*$ across this phase diagram reveals that near $T_c$, the anomalous $dM^*/dT^* > 0$ signature can arise from the high-$T$ behavior of cycloids that is a precursor to the formation of skyrmions.
For $T^* > 1.0$, the magnetization of cycloids at low fields dramatically increases, yielding a substantially positive $dM^*/dT^*$ in the bulk of the cycloid phase.
This behavior is unique to cycloids, as the skyrmion phase maintains $dM^*/dT^* < 0$ at all conditions.
Critically, the $dM^*/dT^* > 0$ signature is a property of the cycloid phase at elevated temperatures, rather than a direct signature of the cycloid-to-skyrmion transition.
This distinction can be seen in Figure \ref{fig:F5}c, which compares the evolution of $dM^*/dT^*$ with the topological index $t$, where $t=0$ for the cycloid and ferromagnet phases and $-1$ for skyrmions.
The region of $dM^*/dT^* > 0$ in the cycloid phase is not accompanied by a change in $t$, indicating that it is a property of the bulk cycloid phase and not a consequence of the cycloid/skyrmion phase transition.
This behavior contrasts with the high-field skyrmion-to-ferromagnet transition, where the $dM^*/dT^* > 0$ feature is centered at the phase boundary and arises due to the phase transition itself.
The evolution of $dM^*/dT^*$ shown in Figure \ref{fig:F5}c is reproduced by trends in microstate fluctuations sampled by Monte Carlo, wherein  $dM^*/dT^*$ is proportional to the covariance of enthalpy and magnetization. 
This covariance term is positive within the bulk of the cycloid phase at finite field, and negative in the skyrmion and ferromagnet phases.
The narrow region of positive $dM^*/dT^*$ near the skyrmion/ferromagnet phase boundary is a signature of the phase transition as the Monte Carlo samples both phases.

At lower temperatures, cycloids are distinguished from skyrmions by the sign of $dM^*/dT^*$, consistent with their behavior in the low-$T$ limit, although the total magnitude of this change is much smaller than the behavior seen near $T_c$.
Notably, for $T^* < 1.0$, there is no temperature dependence in $dM^*/dT^*$, indicating that this positive entropic susceptibility of the cycloid phase is not the explanation for the anomalous low-$T$ feature seen in the \gvse{} \ magnetoentropic map.

The phase diagram of the EA case, representative of \gvs, is shown in Figure \ref{fig:F5}de.
While only the ferromagnetic phase is stable at low $T$, skyrmions and cycloids are stabilized near $T_c$. 
The comparison between the computed phase boundaries and $dM^*/dT^*$ are shown in Figure \ref{fig:F5}de.
An overlay of the topological index $t$ is shown in Figure \ref{fig:F5}f as an unambiguous order parameter for the skyrmion phase.
Similarly to the high-$T^*$ behavior of the EP case, the cycloid phase is characterized by $dM^*/dT^* > 0$, while the skyrmion and ferromagnet phases maintain $dM^*/dT^* < 0$.
In contrast to these features arising from single-phase behavior, the distinctive $dM^*/dT^* \ll 0$ feature at $T^* \approx 1.5$ is aligned with the first-order ferromagnet-to-cycloid and ferromagnet-to-skyrmion transitions, consistent with the discrete change of the magnetic moment at these phase boundaries.

\section{Discussion}

We have demonstrated that magnetoentropic mapping based on high-throughput measurements of $(dM/dT)_H$ can reproduce complex magnetic phase diagrams in uniaxial skyrmion hosts, both identifying the locations of phase boundaries and providing unambiguous signatures of spin-wave single-phase regions.

The most basic features in this data are discontinuities in $M(T)$ associated with phase boundaries between helimagnetic and ferromagnetic states.
These discontinuities appear as peaks and valleys in $(dM/dT)_H$ curves, depending on whether the phase transition is to a phase of higher or lower entropy respectively.
In \gvs{} , the boundaries between the ferromagnet and cycloid/skyrmion phases mapped by $(dM/dT)_H$ show a similar location and spread as those mapped with magnetic susceptibility and SANS \cite{widmann}.
In \gvse{}, the boundary of the ferromagnetic region is clear in our data, but is more difficult to validate as the location of this phase boundary is not well established in the literature due to differences in methods between authors \cite{fujima, bordacs} and variation in crystal quality.

Through the thermodynamic relationship $(dM/dT)_H = (dS/dH)_T$, magnetoentropic data also provides quantitative insight into the properties of single phase regions.
We have shown that close to the Curie temperature $T_c$, and just below the cycloid/skyrmion phase boundary, cycloid phases in both \gvs{} \ and \gvse{} \ are characterized by a $(dM/dT)_H \gg 0$ feature.
This precursor behavior uniquely identifies the cycloid phase in this region of the phase diagram as skyrmions maintain $(dM/dT)_H < 0$ at all conditions.
Furthermore, the magnetoentropic data identifies a new signature of the anomalous low-$T$ phase observed at the cycloid/skyrmion phase boundary in \gvse{} .
This phase had been previously labelled as an anomaly in SANS data \cite{bordacs} and tentatively described as a new spin texture confined to structural domain walls \cite{Geirhos2020}.
We have shown that this phase is characterized by a unique $(dM/dT)_H = (dS/dH)_T \gg 0$ signature that is not explained by any parametrization of the bulk spin Hamiltonian, supporting the proposal that it is controlled by structural domain walls.

The switch in the sign of $(dM/dT)_H$ between the cycloid and skyrmion phases is a particularly valuable practical result of magnetoentropic mapping, as it represents an unambiguous signature distinguishing these otherwise similar phases.
The phase transition between helical/cycloidal order and skyrmions is frequently identified using subtle peaks in AC $(dM/dH)_T$ susceptibility data, leading to controversy in both the location of the phase boundary, and whether it is present at all.
The sign of $(dM/dT)_H$ away from the phase boundary is much easier to measure and interpret, and can be used to confidently map the cycloid/skyrmion boundary.

The physical origin of the $(dM/dT)_H \gg 0$ signature of the cycloid phase is not immediately clear.
Unlike the low-$T$ entropic anomaly in \gvse{} which cannot be reproduced by any parametrization of the bulk crystal spin Hamiltonian, this feature is consistently seen in experimental and simulated derivatives of macroscopic magnetization, as well as fluctuation data obtained from our Monte Carlo sampling.
This result strongly suggests that the $(dM/dT)_H \gg 0$ feature is an intrinsic property of the cycloid phase.
Curiously, this behavior only exists for $T^* > 1.0$, coinciding with the onset of entropically-stabilized skyrmion formation.
This trend suggests the presence of a subtle phase transition near $T^* = 1.0$ within both the cycloid and skyrmion phases, as the low-$T$ enthalpy-dominated behavior is supplanted by entropic stabilization.
While clarifying the underlying physics of this behavior is a promising direction for future theoretical work, the discovery of this subtle behavior supports the unique value of the magnetoentropic mapping technique.

\section{Conclusion}

We demonstrate the utility of the magnetoentropic mapping method in efficiently identifying magnetic phases and determining their boundaries in model uniaxial skyrmion hosts \gvs\ and \gvse. 
When combined with simple computational models of magnetic behavior, magnetoentropic data can distinguish skyrmions from helical/cycloidal spin textures that would otherwise could only be mapped by costly neutron diffraction experiments.
In addition to identifying discontinuities in magnetization analogously to  traditional magnetic susceptibility measurements, magnetoentropic mapping highlights single-phase precursor behavior that can be a much better indicator of the phase transition than the phase boundary itself.
We directly demonstrated this feature in \gvs{} \ by interpreting broadened transitions in structural domains that are not aligned with the applied field, and by characterizing the phase boundaries of a new low-$T$ magnetic phase in \gvse{} .
We conclude that that magnetoentropic mapping is a valuable technique for the rapid characterization of potential uniaxial skyrmion host materials, complementing conventional susceptibility measurements and targeted SANS and LTEM work.

\begin{acknowledgements}
This research was supported by the National Science Foundation (NSF) under DMREF Award No. DMR-1729489. 
Partial support from the NSF Materials Research Science and Engineering Center (MRSEC) at the University of California, Santa Barbara (UCSB), Grant No. DMR-1720256 (IRG-1), is gratefully acknowledged (D.A.K. and A.V.d.V.). 
Use of Shared Experimental Facilities of the UCSB MRSEC (NSF Grant No. DMR 1720256) is acknowledged. 
The UCSB MRSEC is a member of the NSF-supported Materials Research Facilities Network. 
We also acknowledge support from the Center for Scientific Computing (NSF Grant No. DMR-1720256 and NSF Grant No. CNS-1725797), as well as the National Energy Research Scientific Computing Center, a Department of Energy (DOE), Office of Science User Facility supported by DOE Grant No. DE-AC02-05CH11231. 
J.L.Z also acknowledges the support of the NSF Graduate Research Fellowship Program under Grant No. 1650114.
\end{acknowledgements}

\bibliography{GVSSe_citations.bib}

\end{document}